\documentclass[12pt]{article}

\usepackage{amssymb}

\begin{document}

\begin{titlepage}

\begin{flushright}
arXiv:1010.2779
\end{flushright}
\vskip 2.5cm

\begin{center}
{\Large \bf Modeling-Free Bounds on Nonrenormalizable\\
Isotropic Lorentz and CPT Violation in QED}
\end{center}

\vspace{1ex}

\begin{center}
{\large Brett Altschul\footnote{{\tt baltschu@physics.sc.edu}}}

\vspace{5mm}
{\sl Department of Physics and Astronomy} \\
{\sl University of South Carolina} \\
{\sl Columbia, SC 29208} \\
\end{center}

\vspace{2.5ex}

\medskip

\centerline {\bf Abstract}

\bigskip

The strongest bounds on some forms of Lorentz and CPT violation come from
astrophysical data, and placing such bounds may require understanding and modeling
distant sources of radiation. However, it is also desirable to have bounds that do
not rely on these kinds of detailed models.
Bounds that do not rely on any modeling of astrophysical objects
may be derived both from laboratory experiments and certain
kinds of astrophysical observations. The strongest such bounds on
isotropic modifications of electron, positron, and photon dispersion relations
of the form $E^{2}=p^{2}+m^{2}+\epsilon p^{3}$ come from data on cosmological
birefringence, the absence of photon decay, and radiation from lepton beams.
The bounds range in strength from the $4\times10^{-13}$ to $6\times10^{-33}$
$($GeV$)^{-1}$ levels.

\bigskip

\end{titlepage}

\newpage

In recent years, there has grown to be a great deal of interest in the
possibilities of Lorentz and CPT symmetries not being exact in nature.
There is no compelling evidence that calls these symmetries into question, but
if such evidence were uncovered, that would be a discovery of profound importance
and an indication that there are new fundamental physical laws still to be
discovered. Responding to this possibility, there has been an explosion of
experimental work testing Lorentz and CPT invariances. Different types of
experiments, done at many different energy scales, are sensitive to different
possible types of Lorentz violation.

Forms of Lorentz violation whose effects grow more important with
increasing energy present a peculiar case. Obviously, the best bounds on
these forms of Lorentz violation are likely to come from observations of
extremely energetic quanta. The most energetic phenomena we can
study are astrophysical. If we possess an accurate understanding of these
phenomena, we may be able to place very strong constraints. However, it
is not always entirely certain precisely what astrophysical interactions
are responsible for the observations we make here on Earth, and this is a
particular problem at extremely high energies. For example, there is controversy
about the relative importances of inverse Compton scattering and
$\pi^{0}$ decay in the production of observed TeV $\gamma$-rays.

When dealing with the possibility of especially exotic phenomena like
Lorentz violation, it is desirable to have bounds that are as clean as
possible. It is particularly useful to have bounds that require no inferences
about the nature of distant, high-energy emission sources, since the modeling of such
sources could introduce additional uncertainties. There are, however, many
situations in which it is possible to place bounds without any need for such
modeling. For example, bounds may be derived from observations of known particles in
terrestrial laboratory environments.

Of course,
the key distinction is not simply between experiments performed in a laboratory and
observations of astrophysical phenomena. A more important distinction is whether
inferences about
the behavior of distant objects are required. Pulsar wind nebulae like the Crab
nebula are probably
teeming with extremely energetic particles, but drawing inferences about precisely
what is happening inside them---especially at the very highest energies---is
tricky.
Experiments done in the laboratory are much less prone to this kind of uncertainty,
but some astrophysical results are also strongly model independent.
For example, the survival of photons
traveling over long distances is not controversial or subject to modeling
ambiguities. Bounds based simply on the observation of such photons are
extremely clean, and also taking into account their polarizations
will allow us to draw additional conclusions, with only very simple assumptions about
the polarization at the source.

For those dimension-4, renormalizable forms of Lorentz violation that are most
influential in their effects at high energies (meaning those forms of Lorentz violation that add a term $\Delta\propto p^{2}$ to the conventional dispersion
relation $E^{2}=p^{2}+m^{2}$), there has already been a
significant amount of work on both
astrophysical~\cite{ref-stecker,ref-kost11,ref-kost21,ref-altschul7,ref-altschul15}
and
laboratory~\cite{ref-muller3,ref-heckel3,ref-hohensee1,ref-eisele,ref-altschul20,ref-herrmann3,ref-altschul21,ref-bocquet}
bounds. (See~\cite{ref-tables} for a summary of the results.)
The bounds based on
astrophysical data are often stronger, but the laboratory bounds may be more
secure.  However, there has been little corresponding analysis of how
higher-dimensional forms of Lorentz violation may be constrained without using
astrophysical models. In
this paper, we shall perform such an analysis, placing
constraints on dimension-5 Lorentz violations (which add a
$\Delta\propto p^{3}$ to the dispersion relation). Since these
nonrenormalizable forms of Lorentz violation grow very rapidly in importance at
high energies, the best bounds will generally come from observations of very
high-energy quanta---the
large energies serving to enhance ordinarily small effects.

Lorentz violation includes violations of both rotation and boost invariances.
However, we shall concentrate here on forms of Lorentz violation that are purely
isotropic; only boost symmetry is violated. Obviously, perfect isotropy can only
exist in one preferred frame, but if the preferred frame is moving
nonrelativistically relative to the Earth (as is the rest frame of the cosmic
microwave background), anisotropic effects will be suppressed. We shall also
only consider Lorentz violations for electrons, positrons, and photons.

However, even with these simplifying assumptions,
there are multiple forms the Lorentz
violation can take. In the modified dispersion relation,
$E^{2}=m^{2}+p^{2}+\epsilon p^{3}$, the coefficient $\epsilon$ (which parameterizes
the isotropic, dimension-5 Lorentz violation)
may depend on helicity and
particle versus antiparticle identity. Yet not all
such dependences are consistent with the Lorentz violation appearing as part of a
local quantum field theory (QFT)~\cite{ref-myers}, in which the
unperturbed Lagrange density is
${\cal L}_{0}=-\frac{1}{4}F^{\mu\nu}F_{\mu\nu}+\bar{\psi}(i\!\not\!\! D\,\,-m)\psi$.
For photons, a spin-independent
$\epsilon$ is not possible in QFT; the only possible isotropic,
dimension-5 electromagnetic bilinears are equivalent to 
$\Delta{\cal L}_{\gamma}=-\frac{1}{4}(\epsilon_{\gamma}^{+}-\epsilon_{\gamma}^{-})
F^{0\mu}\partial_{0}\tilde{F}_{0\mu}$, where $\tilde{F}$ is the dual of $F$.
For fermions, either a spin-independent
$\Delta{\cal L}_{\psi1}=\frac{1}{8}(\epsilon_{e}^{+}+\epsilon_{e}^{-})
\bar{\psi}\gamma_{0}\partial_{0}^{2}\psi$ or a
spin-dependent
$\Delta{\cal L}_{\psi2}=\frac{1}{8}(\epsilon_{e}^{+}-\epsilon_{e}^{-})
\bar{\psi}\gamma_{0}\gamma_{5}\partial_{0}^{2}\psi$ is
possible. Note that, in addition to having an obvious dependence on particle
chirality, $\Delta{\cal L}_{\psi2}$ produces different dispersion relations
for particles and antiparticles. All the $\epsilon$ coefficients that can exist
in local QFT are odd under CPT.

However, we shall not restrict our attention solely to operators that exist in
QFT. We shall consider more general modified dispersion relations determined by
six parameters,
$\epsilon^{\pm}_{w}$, where $w$ denotes the species ($\gamma$, $e$, or $\bar{e}$)
and the sign label indicates the
helicity. If there is an underlying local field theory, the number of parameters is
only three, since $\epsilon^{+}_{\gamma}=-\epsilon^{-}_{\gamma}$,
$\epsilon_{\bar{e}}^{+}-\epsilon_{\bar{e}}^{-}=\epsilon_{e}^{-}-\epsilon_{e}^{+}$,
and $\epsilon_{\bar{e}}^{+}+\epsilon_{\bar{e}}^{-}=\epsilon_{e}^{+}+
\epsilon_{e}^{-}$. The particular
parameters that can be nonzero in
local QFT are much more interesting than
those that cannot,
since they descend from a fully dynamical theory.
All of the bounds we shall describe will apply equally to electron and positron
coefficients, and we shall denote the coefficient for a generic charged
particle by $\epsilon_{q}^{s}$.

Many of the techniques used to place bounds on
dimension-4 violations of boost invariance can be adapted to constrain
dimension-5 effects as well. However, some tests of boost symmetry only constrain
possible anisotropies in boost behavior, and so they cannot be used to bound the
$\epsilon$ coefficients. The physical effects we shall consider in this paper are
(in order of decreasing sensitivity): photon birefringence, photon decay,
synchrotron radiation, and vacuum Cerenkov (VC) radiation. The impact of
dimension-5 Lorentz violation on certain aspects of these processes has already
been studied in expressly astrophysical contexts~\cite{ref-amelino,ref-schaefer,ref-jacobson1,ref-jacobson2,ref-jacobson4,ref-jacobson3,ref-maccione2}. However,
there has been significant confusion about the rate and spin structure of the photon
decay process, $\gamma\rightarrow e^{-}+e^{+}$, which is important to a number of these studies. The decay rate turns out to be
closely tied to the spins of the decay products; the details of their
relationship---which are also relevant to studies of other kinds of Lorentz
violation---will be discussed for the first time below.

Strong bounds on dimension-5 Lorentz violation in QFT have been derived from
models of the Crab nebula~\cite{ref-maccione2}. While these bounds are stronger than
those derived here, the models involved do not do a perfect job of describing the
Crab's emission spectrum; there is some tension between the populations of high-energy
electrons inferred from the synchrotron and inverse Compton parts of the spectrum.
Consequently, it is
useful to have bounds that do not require any significant degree of
astrophysical modeling. Moreover, the strength of the bounds depends a great deal on
what forms of spin dependence in the Lorentz violation are
considered~\cite{ref-jacobson1}. Strong bounds on some of these dispersion relation
modifications have also
been derived from assumptions of naturalness~\cite{ref-myers}---either assuming that
Lorentz violations in different sectors should be comparable in their orders of
magnitude, because of renormalization group mixing; or assuming that isotropy should
hold only in the rest frame of the cosmic microwave background.

This paper contains a number of new results and places some previously published
results in context. The new elements include:
an analysis of the rate of photon decay in terms of the spin states of the decay
products (a topic that has previously caused some confusion, with different calculations producing different results~\cite{ref-jacobson3}); analysis of
synchrotron
and vacuum Cerenkov processes for accelerator electrons and positrons; and
simultaneous consideration of all possible isotropic, dimension-5 modifications of
particle dispersion relations, whether or not such modifications are allowed in QFT.

The tightest constraint on any $\epsilon$ comes from the polarization of
radiation from
$\gamma$-ray bursts. If $\epsilon_{\gamma}^{+}\neq\epsilon_{\gamma}^{-}$, right-
and left-handed photons have different phase speeds. This leads to birefringence,
a rotation of the polarization over distance. Since the rate of
rotation depends on frequency, an initially linearly polarized polychromatic source
would be depolarized by if the birefringence were strong enough.

There are already several bounds on $\epsilon_{\gamma}^{+}-\epsilon_{\gamma}^{-}$
derived from $\gamma$-ray burst data. We shall quote
the best of these bounds below, but they are
not new to this analysis. However, they are the best limits on
$\epsilon_{\gamma}^{+}-\epsilon_{\gamma}^{-}$ and are included here for two
reasons:  for completeness, and because they will be used
(along with the genuinely new bounds described later in this paper)
to obtain the final, two-sided bounds on some of the other coefficients.

The birefringence effect has been considered in several papers.
In~\cite{ref-jacobson4}, the reported polarization
$\Pi_{021206}=(80\pm20)$\%~\cite{ref-coburn} of the
$\gamma$-ray burst GRB 021206 ($\Pi$ denoting the polarization
fraction of the $\gamma$-rays, averaged over the observations' energy range)
for photons covering a 15--2000 keV range
was used as the basis for a bound on $\epsilon_{\gamma}^{+}-\epsilon_{\gamma}^{-}$.
However, the accuracy of the polarization measurement has been called strongly
into question~\cite{ref-rutledge,ref-wigger}.
In~\cite{ref-kost23} two more polarized $\gamma$-ray
bursts, GRB~930131 and GRB~960924, with observed polarizations of
$\Pi_{930131}>35$\% and $\Pi_{960924}>50$\% were considered. 
These three sources are fairly distant, each with redshift $z\gtrsim0.1$.
Most recently, the polarization of GRB 041219a (a source with a
pseudo-redshift $z\gtrsim0.2$) over an energy range of 100--350
keV~\cite{ref-mcglynn}, has been used to derive another bound~\cite{ref-stecker2}.

With the simplest assumption about the initial polarization of a $\gamma$-ray burst,
that the radiation began completely linearly polarized, the survival of
the observed degrees of polarization over cosmological distances indicates that
$(\epsilon_{\gamma}^{+}-\epsilon_{\gamma}^{-})\lesssim[d
(\omega_{2}^{2}-\omega_{1}^{2})]^{-1}$, where $d$ is the distance to the source,
while $\omega_{1}$ and $\omega_{2}$ are the lower and upper limits of the observed
$\gamma$-ray frequency range. That the inequality is satisfied ensures that the
plane of polarization does not change by an ${\cal O}(1)$ angle over the energy
range involved. The best published bound resulting from this kind of analysis comes from~\cite{ref-kost23}; in the notation of that paper, the bound is
\begin{equation}
\label{eq-kV}
\left|k_{(V)00}^{(5)}\right|=\left|-\sqrt{\pi}(\epsilon_{\gamma}^{+}-
\epsilon_{\gamma}^{-})\right|<10^{-32}\,({\rm GeV})^{-1}.
\end{equation}
The details of the analysis are found
in~\cite{ref-jacobson4,ref-kost23}; we omit further
discussion of how they were derived, because no modifications to the analysis are
necessary for the present purposes.

The order of magnitude of the bound (\ref{eq-kV}) is fairly robust, because the only
assumptions it requires are related to the sources' distances
(for which lower limits are available) and the initial polarization of the
radiation. Barring some exceedingly unlikely
conspiracy of circumstance, in which the radiation began with a
frequency-dependent plane of polarization, such that, after birefringence, the
polarizations all lined up at the time of observation, a strong initial linear
polarization represents the most natural hypothesis.

Moreover, even if there were reason to doubt the accuracy of the bound
(\ref{eq-kV}) derived in~\cite{ref-kost23}, the constraint is so strong that, even if
the constraint on $(\epsilon_{\gamma}^{+}-\epsilon_{\gamma}^{-})$ were actually
determined to be $10^{11}$ times weaker, none of the later results in this paper
would be affected. In fact, if the GRB 021206 results were accurate, the bound would
actually be about a factor of 7 stronger. Moreover, the
observed polarization of synchrotron x-rays from the Crab nebula
indicates a bound on $|\epsilon_{\gamma}^{+}-\epsilon_{\gamma}^{-}|$
at the $10^{-27}$ (GeV)$^{-1}$ level or better~\cite{ref-maccione1}.
Henceforth, we shall simply neglect photon birefringence, because it is so much
more strongly constrained than any of the other effects to be considered. Including
the possibility of birefringence at the allowed level (\ref{eq-kV}) would not affect
any of our other numerical results. So we shall only consider the possibility of a
single $\epsilon_{\gamma}=\epsilon_{\gamma}^{+}=\epsilon_{\gamma}^{-}$.
(Note that in the QFT framework, in which $\epsilon_{\gamma}^{-}=-\epsilon_{\gamma}^{+}$, this means neglecting Lorentz
violation in the photon sector altogether; there can be no dimension-5 photon Lorentz
violation in QFT without birefringence.)

There are no other
existing bounds that can so simply be incorporated into this analysis without
modification. The next process to consider is photon decay,
$\gamma\rightarrow e^{-}+e^{+}$, which is ordinarily forbidden by boost invariance
and momentum conservation. However, it has already been noted that the existence of
Lorentz violation may make this decay allowed for sufficiently energetic photons.
Most previous discussions have focused primarily on the threshold for the process;
however, understanding the process's rate---especially how the rate depends on the
spin states of the decay products---is also crucial to understanding the bounds
that follow from the observed absence of the decay. Some correct bounds have
already been published~\cite{ref-jacobson4}, but there has been significant
confusion about the spin structure of the decay.

It is not necessary that the photon decay threshold correspond to a
symmetric configuration in
which the daughter particles are emitted with equal energies. In fact,
even if $\epsilon_{e}^{s}=\epsilon_{\bar{e}}^{s'}$, the threshold (at which
all the particles involved are moving collinearly) may be
asymmetric; but nonetheless we shall focus on the symmetric threshold, which is
simpler. In some regimes, including the
possibility of an asymmetric threshold produces a modest improvement in the numerical
bounds that can be set. However, there is no such improvement possible in the
final constraints to be derived in this paper.

To get the most comprehensive bounds possible from a study of photon decay, we must
look at the spin structure of the reaction and look at energies above the threshold,
so that the particles involved need not be perfectly collinear.
We shall discuss configurations near the symmetric threshold, but the asymmetric case
is similar. If the $e^{-}$ and $e^{+}$ each make an angle $\theta$ with the
original photon's momentum $\vec{p}$, energy-momentum conservation requires
\begin{equation}
\label{eq-theta}
\epsilon_{\gamma}-\frac{1}{4}(\epsilon_{e}^{s}+\epsilon_{\bar{e}}^{s'})\sec^{3}\theta=\frac{4m^{2}+p^{2}\tan^{2}\theta}{p^{3}}.
\end{equation}
If $\theta\lesssim\frac{m}{p}\ll 1$, the right-hand side of (\ref{eq-theta}) is only
modestly greater than $\frac{4m^{2}}{p^{3}}$. The characteristic range of angles
$\theta$ for decays only slightly above threshold is therefore
$\theta\lesssim\frac{m}{p}$.
For these angles, $\sec\theta\approx1$, and so the left-hand side of
(\ref{eq-theta})---which is the combination of Lorentz violation coefficients upon
which the decay rate depends---reduces to the $\theta$-independent sum
$\epsilon_{\gamma}-\frac{1}{4}\epsilon_{e}^{s}-\frac{1}{4}\epsilon_{\bar{e}}^{s'}$.

Since the initial photon is a $J=1$ state, the final spins must be parallel,
oriented the same way along the photon's propagation axis. At
threshold, where all the particles are collinear, this means the $e^{+}$ and
$e^{-}$ must have the same helicity as the photon. However, if the particles each veer
off at an angle
$\theta$, each may be found in a helicity state opposite that of the initial photon
with probability $\sin^{2}\frac{\theta}{2}$. The rate for the decay process in which
both final particles have the same helicity as the photon is 
$\Gamma_{0}\sim e^{2}p\cos^{4}\frac{\theta}{2}$. For each particle with helicity
opposite that of the photon, one factor of $\cos^{2}\frac{\theta}{2}$ changes to
$\sin^{2}\frac{\theta}{2}$. These results are consistent with the very different
energy dependences inferred for the decay rates by Jacobson, Liberati, and Mattingly
in~\cite{ref-jacobson2} and~\cite{ref-jacobson3}. They identified
$\Gamma\sim e^{2}p$
if the final spin states are arbitrary and the $e^{-}$ and $e^{+}$ dispersion
relations are independent of spin; contrariwise, if the daughter particles have
opposite spins but
still the same dispersion relation, $\Gamma\sim e^{2}\epsilon p^{2}$.
The difference derives from the trigonometric 
factors in the decay
rates. If the photon has helicity $s$, each daughter particle with helicity $-s$
slows the decay by a factor of $\tan^{2}\frac{\theta}{2}$; and near threshold,
$\tan^{2}\frac{\theta}{2}\sim\frac{m^{2}}{p^{2}}\sim\epsilon p$.

If the slowest possible decay rate, $\Gamma_{0}\tan^{4}\frac{\theta}{2}$, is large
compared with the reciprocal of a
photon's time of flight, the reaction $\gamma\rightarrow e^{-}+e^{+}$ has time to
occur for any of the four possible outgoing helicity states. For extremely
energetic $\gamma$-rays
$\Gamma_{0}\sim10^{25}\left(\frac{p}{{10\,{\rm TeV}}}\right)$.
For a 50 TeV photon, $\frac{m}{p}=10^{-8}$; this makes the typical angle for a
near-threshold decay $\theta\sim10^{-8}$. The decay lifetime of such a 50 TeV photon
is therefore $16\Gamma_{0}^{-1}\theta^{-4}\sim10^{8}$ s. For a source such as the
Crab nebula, lying approximately 2 kpc away and emitting photons above 50 TeV, the
time of flight is $2\times10^{11}$ s; such photons have ample time to decay, provided
the condition,
$4\epsilon_{\gamma}-\epsilon_{e}^{s}-\epsilon_{\bar{e}}^{s'}>\frac{20m^{2}}{p^{3}}=
4\times10^{-20}$ $($GeV$)^{-1}$ is met. Conversely, the stability of the 50 TeV
photons indicates that
\begin{equation}
\label{eq-decay}
4\epsilon_{\gamma}-\epsilon_{e}^{s}-\epsilon_{\bar{e}}^{s'}<4\times10^{-20}\,
({\rm GeV})^{-1};
\end{equation}
our analysis of the decay products' spin behavior enables us to set this bound
for all combinations of $s$ and $s'$. 

The derivation of the result (\ref{eq-decay}) required
no inferences about how the TeV $\gamma$-rays were produced at
their source; it only matters that the photons traversed the full
distance from the Crab nebula to Earth. This independence of source characteristics
is a general feature of bounds inferred from the absence of photon decay.
The bounds (\ref{eq-decay}) are the strongest ones available on
the fermionic $\epsilon_{q}^{s}$ coefficients that are independent of a model of the
source, but they have the disadvantage of being strictly one-sided.

It is possible to place two-sided bounds using observations of the synchrotron losses at the Large Electron-Positron Collider (LEP).
This has already been done for the dimension-4 coefficients in the
standard model extension~\cite{ref-altschul20,ref-altschul5}.
Yet in a theory that lacks Lorentz invariance
and is not necessarily based on QFT, it
may not be immediately obvious what form the electromagnetic coupling
to relativistic charged matter should take.
We shall assume that the modified theory has a gauge-invariant single-particle
Hamiltonian $H$ for each charge, and this $H$ is (almost) minimally coupled to the
electromagnetic field. Then $H$ is a function of the electromagnetic field
(almost) solely via the combination
$\vec{p}-e\vec{A}$. It follows that the electromagnetic field couples
principally to the
four-velocity $v^{\mu}$. If the electromagnetic sector were completely conventional,
then given a known particle trajectory, we could determine the emitted radiation by
standard methods.
Moreover, even if the photon dispersion relation is also Lorentz violating,
$\epsilon_{\gamma}$ has minimal impact on the synchrotron process, because the
individual photons that are emitted have energies much smaller than do the orbiting
charges.

So neglecting $\epsilon_{\gamma}$, the power
radiated by a synchrotron electron is $P=\frac{e^{2}a^{2}}{6\pi m^{2}}\gamma^{4}$,
where $a$ is the magnitude of the acceleration and $\gamma=(1-v^{2})^{-1/2}$ is
the usual Lorentz factor. For ultrarelativistic particles,
$\gamma$ is a very rapidly increasing function of $v$.
A charged particle's velocity in the presence of the Lorentz violation is
\begin{equation}
v=\frac{\partial E}{\partial p}\approx1-\frac{m^{2}}{2p^{2}}+\epsilon^{s}_{q}p
\end{equation}
at high energies. This makes the radiated power
\begin{equation}
\label{eq-P}
P=P_{0}(1+4\gamma^{2}\epsilon^{s}_{q}p),
\end{equation}
where $P_{0}$ is the radiation rate in the conventional case, and
$\epsilon_{q}^{s}$
is whichever coefficient is relevant for the charged particle in question.

A precise determination of the beam energy $E_{b}$ at LEP was important, since the
machine was used to make precision W and Z boson mass measurements.
The primary method of determining $E_{b}$ involved measuring the magnetic field
profile with nuclear magnetic resonance and studying the beam
trajectory. The field strength and the trajectory through the bending magnets were
known to high precision, and together those quantities determined $E_{b}$.

The synchrotron radiation rate was used as part of a redundant measurement of the
beam energies. This redundant measurement looked at the synchrotron tune, $Q_{s}$,
which is the ratio of the synchrotron oscillation frequency to the
orbital frequency~\cite{ref-assmann}. Synchrotron oscillations occur when beam
particles' energies differ from the nominal value $E_{b}$. Less energetic particles revolve around smaller orbits and thus travel between the accelerating
radio
frequency (RF) cavities more quickly. They arrive at the cavities earlier in the
RF cycle and receive larger than nominal energy boosts, pushing their energies
back towards $E_{b}$. The opposite effect occurs for particles with greater than the
nominal energy.

$Q_{s}$ is proportional to $(P/E_{b})^{2}$, and since $E_{b}$ is known from other
measurements, synchrotron tune measurements can be
turned around to give measurements of the $e^{-}$ and $e^{+}$ radiation rates
(and hence velocities). The extent to which the synchrotron tune energy measurement
agreed with the primary energy calibration determines how small $\epsilon$ must be.
A conservative $2\sigma$ bound on the fractional deviation of $P$ from its
conventionally expected value is $\left|\frac{\Delta P}{P}\right|<6\times 10^{-4}$,
for measurements performed at the Z pole energy of $E_{b}=91$ GeV.

The LEP beams were maintained in transverse polarization states, except sometimes
near interaction points. (An initially pure helicity state would precess many times
during a single orbit.) Since the average beam helicity was zero, the synchrotron
radiation rate would depend only on the average
$\frac{1}{2}(\epsilon_{q}^{+}+\epsilon_{q}^{-})$ for each beam. Moreover, the
synchrotron oscillations could not be observed separately for the
counterpropagating $e^{-}$ and $e^{+}$ beams; the oscillations were only
observed for the beam system as a whole. With nominally equal $e^{-}$ and
$e^{+}$ beam currents, any effect to be observed would
have been averaged over the particle types. (The existence of differing
$Q_{s}$ values for $e^{-}$ and $e^{+}$ is another phenomenon that does not
occur in QFT.) The agreement of
the synchrotron tune measurement with expectations therefore yields the constraint
\begin{equation}
\label{eq-sync}
|\epsilon^{+}_{e}+\epsilon^{-}_{e}+\epsilon^{+}_{\bar{e}}+\epsilon^{-}_{\bar{e}}|<
\frac{|\Delta P/P|}{\gamma^{2}p}=2.2\times10^{-16}\,({\rm GeV})^{-1},
\end{equation}
using $p\approx E_{b}$.
This is weaker than the photon survival bound, but it is two sided.

In order to have two-sided constraints on all the coefficients, one more process
needs to be considered.
Photon decay would occur if photon energies grew more rapidly with momentum than electron and positron energies; conversely, if the fermion energies were
more rapidly increasing functions of $p$, the VC radiation
process $e^{\pm}\rightarrow e^{\pm}+\gamma$ could occur. However, the data from LEP
showed no evidence of this process up to $E_{b}=104.5$ GeV. Above threshold, such
radiation would be quite rapid, so any combination of Lorentz violation
coefficients that would allow the radiation to occur can be reliably ruled out.

There are two type of VC radiation: hard and soft. The threshold for
emission of a soft photon is basically insensitive to $\epsilon_{\gamma}$, since
the photon energy is so small. The threshold in this case is
$p_{T}=(m^{2}/2\epsilon_{q}^{s})^{1/3}$ (which is the momentum at which a charge's
speed $v$ exceeds 1), and the
absence of soft emission by a charged particle of momentum $p\approx E_{b}$ implies
\begin{equation}
\label{eq-softthreshold}
\epsilon_{q}^{s}<\frac{m^{2}}{2p^{3}}=1.5\times10^{-13}.
\end{equation}
This bound applies to both helicity states of both electrons and positrons.

In a hard VC process, in which a charge of momentum $p$ retains a portion
$xp$ of its momentum while emitting a collinear photon of momentum $(1-x)p$, the
energy-momentum conservation condition reduces to
$\epsilon_{\gamma}-\left(\frac{1+x}{1-x}\right)\epsilon_{q}^{s}=-\frac{1}{x(1-x)}
\frac{m^{2}}{p^{3}}$. (This assumes there is no change in the charge's helicity
state, but including helicity-changing reactions only produces additional, more
complicated inequalities that do not improve the ultimate numerical results.)

The absence of VC radiation from charges of momentum $p$
implies the threshold condition cannot be satisfied  for any relevant $x$, and
thus
\begin{equation}
\label{eq-threshold}
\epsilon_{\gamma}-\left(\frac{1+x}{1-x}\right)\epsilon_{q}^{s}>-\frac{1}{x(1-x)}
\frac{m^{2}}{p^{3}}
\end{equation}
for all $\frac{m}{p}\ll x<1$. The lower limit on $x$ comes from the requirement
that the charge remain ultrarelativistic after radiating, or $xp\gg m$.
The inequality (\ref{eq-threshold}) is needed to place a lower bound on
$\epsilon_{\gamma}$. The bound becomes more sensitive to $\epsilon_{\gamma}$
(as opposed to $\epsilon_{q}^{s}$) for smaller $x$, but the magnitude of the
right-hand side of (\ref{eq-threshold}) grows rapidly as $x\rightarrow0$.
(Note that the soft threshold corresponds to $x\approx1$.)
Although (\ref{eq-threshold}) represents a continuous set of inequalities, the
key features are captured by the $x=\frac{1}{2}$ version,
\begin{equation}
\epsilon_{\gamma}-3\epsilon_{q}^{s}>-\frac{4m^{2}}{p^{3}}=-3\times10^{-13}\,
({\rm GeV})^{-1},
\end{equation}
again using $p\approx E_{b}$.

Separate bounds on important combinations of $\epsilon$ coefficients can
be extracted from (\ref{eq-kV}), (\ref{eq-decay}), (\ref{eq-sync}),
(\ref{eq-softthreshold}) and
(\ref{eq-threshold}) by linear programming. The results are given in
table~\ref{table-separate}, which contains entries with each of of the
characteristic strengths seen in the raw bounds. (A bound at a given level
typically depends on a raw bound at the same level, as well as the stronger
raw bounds discussed previous to it in this paper.
A result at the $10^{-16}$ level, for example, relies on the
birefringence, photon decay, and synchrotron data.) Note that, since all the
raw results treat electrons and positrons, of positive and negative helicity,
equivalently, the constraints on $\epsilon_{e}^{+}$, $\epsilon_{e}^{-}$,
$\epsilon_{\bar{e}}^{+}$, and $\epsilon_{\bar{e}}^{-}$ are all identical.
Table~\ref{table-separate} also includes bounds for the QFT case, in which
there are only three independent coefficients. In that case, the bounds are
generally better, and the weakest raw constraints
(coming from the absence of VC radiation at LEP) are not needed.

\begin{table}
\begin{center}
\begin{tabular}{|c|c|c|c|c|}
\hline
Coefficient & Maximum & Minimum & QFT Maximum & QFT Minimum\\
\hline
$\epsilon_{\gamma}^{+}-\epsilon_{\gamma}^{-}$ & $6\times10^{-33}$
& $-6\times10^{-33}$ & $6\times10^{-33}$ & $-6\times10^{-33}$ \\
$\epsilon_{\gamma}^{+}+\epsilon_{\gamma}^{-}$ & $6\times10^{-17}$ &
$-6\times10^{-13}$ & $-$ & $-$ \\
$\epsilon_{e}^{+}$, $\epsilon_{e}^{-}$, $\epsilon_{\bar{e}}^{+}$, and 
$\epsilon_{\bar{e}}^{-}$ & $10^{-13}$ & $-3\times10^{-13}$ & $1.1\times10^{-16}$ &
$-2\times10^{-20}$ \\
$\epsilon_{e}^{+}+\epsilon_{e}^{-}$
& $2\times10^{-13}$ & $-2\times10^{-13}$ & $1.1\times10^{-16}$ &
$-4\times10^{-20}$ \\
$\epsilon_{e}^{+}-\epsilon_{e}^{-}$ & $4\times10^{-13}$ & $-4\times10^{-13}$ &
$1.1\times10^{-16}$ & $-1.1\times10^{-16}$ \\
\hline
\end{tabular}
\caption{
\label{table-separate}
Maximum and minimum allowed values of electron, positron, and photon $\epsilon$
coefficients, in units
of (GeV)$^{-1}$. The fourth and fifth columns give the bounds if only the
coefficients that can exist in local QFT are considered. The dashes in those
columns denote combinations that are identically zero in the QFT framework.}
\end{center}
\end{table}

For several of the $\epsilon$ coefficients, the results given in 
Table~\ref{table-separate} are the first modeling-free constraints to be placed
on them: none of these bounds require any understanding of the behavior of
high-energy astrophysical sources, as opposed to many earlier bounds, which
relied on our having an accurate picture of how the photons we see coming
from particular sources are produced. So far, these results are limited to
isotropic
forms of Lorentz violation, but many of the same techniques discussed here could
also be used to study anisotropic, dimension-5 Lorentz- and CPT-violating
operators (which are extremely numerous). Moreover, the results are not limited to
those forms of Lorentz violation that can exist in QFT; purely phenomenalistic
modifications of particle dispersion relations---although they are
less plausible---were also found to be subject to strong bounds.

\end{document}